\documentclass[
reprint,
superscriptaddress,
showpacs,preprintnumbers,
amsmath,amssymb,
aps,
prc,
]{revtex4-1}

\usepackage{graphicx}
\usepackage{dcolumn}
\usepackage{multirow}
\usepackage{bm}

\begin{document}


\title{High Precision Measurement of the $^{19}$Ne Half-life using real-time digital acquisition} 

\author{C. Fontbonne}
\affiliation{Normandie Univ, ENSICAEN, UNICAEN, CNRS/IN2P3, LPC Caen, 14000 Caen, France}
\author{P. Uji\'{c}}
\affiliation{Vinca Institute of Nuclear Sciences, University of Belgrade, P.O.Box 522, 11070 Belgrade, Serbia}
\author{F. de Oliveira Santos}
\affiliation{GANIL, CEA/DRF-CNRS/IN2P3, Bvd Henri Becquerel, 14076 Caen, France}
\author{X. Fl\'echard}
\email{flechard@lpccaen.in2p3.fr}
\affiliation{Normandie Univ, ENSICAEN, UNICAEN, CNRS/IN2P3, LPC Caen, 14000 Caen, France}
\author{F. Rotaru}
\affiliation{Horia Hulubei National Institute for Physics and Nuclear Engineering, P.O. Box MG-6, 76900 Bucharest, Romania}
\author{N. L. Achouri}
\affiliation{Normandie Univ, ENSICAEN, UNICAEN, CNRS/IN2P3, LPC Caen, 14000 Caen, France}
\author{V. Girard Alcindor}
\affiliation{GANIL, CEA/DRF-CNRS/IN2P3, Bvd Henri Becquerel, 14076 Caen, France}
\author{B. Bastin}
\affiliation{GANIL, CEA/DRF-CNRS/IN2P3, Bvd Henri Becquerel, 14076 Caen, France}
\author{F. Boulay}
\affiliation{GANIL, CEA/DRF-CNRS/IN2P3, Bvd Henri Becquerel, 14076 Caen, France}
\author{J. B. Briand}
\affiliation{GANIL, CEA/DRF-CNRS/IN2P3, Bvd Henri Becquerel, 14076 Caen, France}
\author{A. M. S\'anchez-Ben\'{\i}tez}
\affiliation{Departamento de Ciencias Integradas, University of Huelva, 21071 Huelva, Spain}
\affiliation{GANIL, CEA/DRF-CNRS/IN2P3, Bvd Henri Becquerel, 14076 Caen, France}
\author{H. Bouzomita}
\affiliation{GANIL, CEA/DRF-CNRS/IN2P3, Bvd Henri Becquerel, 14076 Caen, France}
\author{C. Borcea}
\affiliation{Horia Hulubei National Institute for Physics and Nuclear Engineering, P.O. Box MG-6, 76900 Bucharest, Romania}
\author{R. Borcea}
\affiliation{Horia Hulubei National Institute for Physics and Nuclear Engineering, P.O. Box MG-6, 76900 Bucharest, Romania}
\author{B. Blank}
\affiliation{CENBG, 19 Chemin du Solarium, CS 10120, F-33175 Gradignan Cedex, France }
\author{B. Carniol}
\affiliation{Normandie Univ, ENSICAEN, UNICAEN, CNRS/IN2P3, LPC Caen, 14000 Caen, France}
\author{I. \v{C}elikovi\'{c}}
\affiliation{Vinca Institute of Nuclear Sciences, University of Belgrade, P.O.Box 522, 11070 Belgrade, Serbia}
\author{P. Delahaye}
\affiliation{GANIL, CEA/DRF-CNRS/IN2P3, Bvd Henri Becquerel, 14076 Caen, France}
\author{F. Delaunay}
\affiliation{Normandie Univ, ENSICAEN, UNICAEN, CNRS/IN2P3, LPC Caen, 14000 Caen, France}
\author{D. Etasse}
\affiliation{Normandie Univ, ENSICAEN, UNICAEN, CNRS/IN2P3, LPC Caen, 14000 Caen, France}
\author{G. Fremont}
\affiliation{GANIL, CEA/DRF-CNRS/IN2P3, Bvd Henri Becquerel, 14076 Caen, France}
\author{G. de France}
\affiliation{GANIL, CEA/DRF-CNRS/IN2P3, Bvd Henri Becquerel, 14076 Caen, France}
\author{J. M. Fontbonne}
\affiliation{Normandie Univ, ENSICAEN, UNICAEN, CNRS/IN2P3, LPC Caen, 14000 Caen, France}
\author{G.F. Grinyer}
\affiliation{GANIL, CEA/DRF-CNRS/IN2P3, Bvd Henri Becquerel, 14076 Caen, France}
\affiliation{Department of Physics, University of Regina, Regina, SK S4S 0A2, Canada}
\author{J. Harang}
\affiliation{Normandie Univ, ENSICAEN, UNICAEN, CNRS/IN2P3, LPC Caen, 14000 Caen, France}
\author{J. Hommet}
\affiliation{Normandie Univ, ENSICAEN, UNICAEN, CNRS/IN2P3, LPC Caen, 14000 Caen, France}
\author{A. Jevremovi\'{c}}
\affiliation{Vinca Institute of Nuclear Sciences, University of Belgrade, P.O.Box 522, 11070 Belgrade, Serbia}
\author{M. Lewitowicz}
\affiliation{GANIL, CEA/DRF-CNRS/IN2P3, Bvd Henri Becquerel, 14076 Caen, France}
\author{I. Martel}
\affiliation{Departamento de Ciencias Integradas, University of Huelva, 21071 Huelva, Spain}
\author{J. Mrazek}
\affiliation{Nuclear Physics Institute, ASCR CZ-25068, \v{R}e\v{z}, Czech Republic}
\author{M. Parlog}
\affiliation{Normandie Univ, ENSICAEN, UNICAEN, CNRS/IN2P3, LPC Caen, 14000 Caen, France}
\affiliation{Horia Hulubei National Institute for Physics and Nuclear Engineering, P.O. Box MG-6, 76900 Bucharest, Romania}
\author{J. Poincheval}
\affiliation{Normandie Univ, ENSICAEN, UNICAEN, CNRS/IN2P3, LPC Caen, 14000 Caen, France}
\author{D. Ramos}
\affiliation{Universidade de Santiago de Compostela, E-15706 Santiago de Compostela, Spain}
\author{C. Spitaels}
\affiliation{GANIL, CEA/DRF-CNRS/IN2P3, Bvd Henri Becquerel, 14076 Caen, France}
\author{M. Stanoiu}
\affiliation{Horia Hulubei National Institute for Physics and Nuclear Engineering, P.O. Box MG-6, 76900 Bucharest, Romania}
\author{J. C. Thomas}
\affiliation{GANIL, CEA/DRF-CNRS/IN2P3, Bvd Henri Becquerel, 14076 Caen, France}
\author{D. Toprek}
\affiliation{Vinca Institute of Nuclear Sciences, University of Belgrade, P.O.Box 522, 11070 Belgrade, Serbia}

\date{\today}

\begin{abstract}
The half-life of $^{19}$Ne has been measured using a real-time digital multiparametric acquisition system providing an accurate time-stamp and relevant information on the detectors signals for each decay event. An exhaustive offline analysis of the data gave unique access to experimental effects potentially biasing the measurement. After establishing the influence factors impacting the measurement such as after-pulses, pile-up, gain and base line fluctuations, their effects were accurately estimated and the event selection optimized. The resulting half-life, $17.2569\pm0.0019_{(stat)}\pm0.0009_{(syst)}$~s, is the most precise up to now for $^{19}$Ne. It is found in agreement with two recent precise measurements and not consistent with the most recent one [L.J. Broussard {\it et al.}, Phys. Rev. Lett. {\bf112}, 212301 (2014)] by 3.0 standard deviations. The full potential of the technique for nuclei with half-lives of a few seconds is discussed.
\end{abstract}


\maketitle

\section{Introduction.} 
Nuclear $\beta$ decay offers an attractive electroweak process for the study of fundamental interactions~\cite{Holstein14,Severijns14} and plays an important role in nucleosynthesis, particularly in $\it{r}$-process nucleosynthesis~\cite{Arnould2007} where the half-life of exotic $\beta$ emitters is often a key parameter. Precise half-life measurements are also important in the investigation of the influence of electron screening on radioactive $\beta$ decays~\cite{Ujic2013}. During the last decade the main motivation for accurate half-life measurements of $\beta$ emitters has been related to the determination of the up-down element $V_{ud}$ of the Cabibbo-Kobayashi-Maskawa (CKM) \cite{Cabibbo, Kobayashi-Maskawa} quark mixing matrix. The unitarity condition of the CKM mixing matrix is one of the sensitive probes to test the consistency of the three generation standard electroweak model, and any deviation from unitarity would imply new physics. The precision of this test is presently limited by the uncertainty on the value of the dominant element, $V_{ud}$. To date, the most precise value of $V_{ud}$ is inferred from the so-called $ft-$values (the product of the statistical rate function $f$ of the transition with the partial half-life $t$ of the decaying state) of the superallowed $0^+ \rightarrow 0^+$ pure Fermi $\beta$ transitions of 14 nuclei, yielding $V_{ud}=0.97417 \pm 0.00021$~\cite{Hardy_Towner_2015}. This analysis requires, for each nucleus, to combine high precision measurements of its $Q-$value, of the $0^+ \rightarrow 0^+$ branching ratio, and of its half-life.

 The mixed Fermi and Gamow-Teller transitions of $T=1/2$ mirror $\beta$ decays provide the second most precise value of $V_{ud}$ \cite{Grinyer2015,Fenker2017}. Transitions between mirror nuclei, like in neutron decay, are induced by both the vector and axial-vector interactions and the extraction of $V_{ud}$ is conducted in analogy with free neutron decay. While mirror transitions offer a complementary and promising set of nuclei for the extraction of $V_{ud}$, they also require the additional measurement of the Fermi fractions of the transitions~\cite{Naviliat-Cuncic2009}. These are inferred from correlation measurements for a set of five $T=1/2$ nuclear mirror transitions, leading today to $V_{ud}=0.9728 \pm 0.0014$~\cite{Fenker2017}. 
The $^{19}$Ne and $^{37}$K nuclei hold a special place in this analysis as their decays yield the most precise contributions in the set of $T=1/2$ nuclear mirror transitions. The uncertainty on $V_{ud}$ is still dominated by the precision of the Fermi fractions measurements~\cite{Naviliat-Cuncic2009}, but the question is raised about the reliability of the previous half-life estimation of $^{19}$Ne. This is due to significant discrepancies of more than two standard deviations between the three most recent and most precise measurements: $17.262 \pm 0.007$~s \cite{Triambak2012}, $17.254 \pm 0.005$~s \cite{Ujic2013} and $17.2832 \pm 0.0051_{(stat)} \pm 0.0066_{(syst)}$~s \cite{Broussard2014}.

The present work reports a new high statistics measurement of the $^{19}$Ne half-life whose main goal is to provide a clarification of the situation regarding the previous estimates. We also intend to demonstrate the capabilities of modern and fast real-time digital acquisition techniques, not yet commonly used in nuclear experimental physics. To our knowledge, such acquisition systems with an effective dead time below 1~$\mu$s have been used only twice for nuclear half-life measurements~\cite{Flechard2010,Naviliat2017}. We show here that, with appropriate precautions, relative uncertainties of $10^{-4}$ and below can be easily achieved for half-life measurements of nuclei with production rates on the order of $10^5$~pps. This opens up new perspectives in a context where large efforts are dedicated to improved precision in half-life measurements~\cite{Triambak2012, Ujic2013, Broussard2014, Grinyer2015, Bacquias2012, Shidling2014, Grinyer2015_b} and Fermi fraction measurements~\cite{Ban2013,Fenker2017} to extract $V_{ud}$ from mirror transitions. Our analysis also shows evidence for sources of systematic errors at the level of $10^{-4}$ in relative precision that $must$ be addressed using a multi-parametric acquisition system.
\section{Experimental set-up}
The $^{19}$Ne beam was produced at the SPIRAL1 facility of GANIL (Caen, France) by impinging $^{20}$Ne$^{10+}$ ions at 95~MeV/u on a graphite target. The $^{19}$Ne$^{3+}$ ions produced within the target-ECR source were post-accelerated by the CIME cyclotron at E$_{lab}$=5.0982 MeV/u and sent to the G2 experimental hall. The maximum beam intensity was $5\times 10^{6}$ pps. In order to eliminate potential beam impurities from $^{19}$O, a carbon stripper was introduced at the exit of the cyclotron and the ion optics downstream the stripper was set to select the charge-state 10+, larger than the maximum possible charge state of $^{19}$O. The relative contamination due to $^{19}$O was then measured using a silicon detector. Thanks to this effective method, this relative contamination was found to be below $10^{-5}$.
\par
The beam was periodically implanted in a $20\times20$~mm$^2$ 50~$\mu$m thick lead target. According to simulations, the implantation depth within the target was between 25 and 30~$\mu$m. With the reasonable assumption that the diffusion coefficient of Ne in Pb is comparable to those measured for Ne in Ag~\cite{Glyde1967} and He in Au~\cite{Sciani1983}, the loss of Ne atoms due to diffusion out of the target was found completely negligible, with a maximum bias on the half-life estimated below $10^{-20}$ s.
 A typical half-life measurement cycle consisted of 6.4~s of beam implantation followed by 440~s of $\beta$ decay data taking. Between these two periods, the lead target was alternately moved in front of the detection system or in the beam axis using a rotating arm. The beam was on only during the implantation period. The short duration of the implantation period, about a third of the $^{19}$Ne half-life, was chosen to limit the maximum count rate of the detectors below 2$\times10^5$ counts per second. The long duration of $\beta$ decay data taking period, about 25 times the $^{19}$Ne half-life, ensured a precise control of the background for each cycle. The detection system consisted of a 33$\times$33~mm$^2$ 5~mm thick BC-400 plastic scintillator located $\sim$5~mm away from the lead target. As shown in figure 1, the light from the scintillator was collected using two light guides connected to two photomultipliers (PM) R2248 from Hamamatsu.
\begin{figure}
\includegraphics[width=85mm]{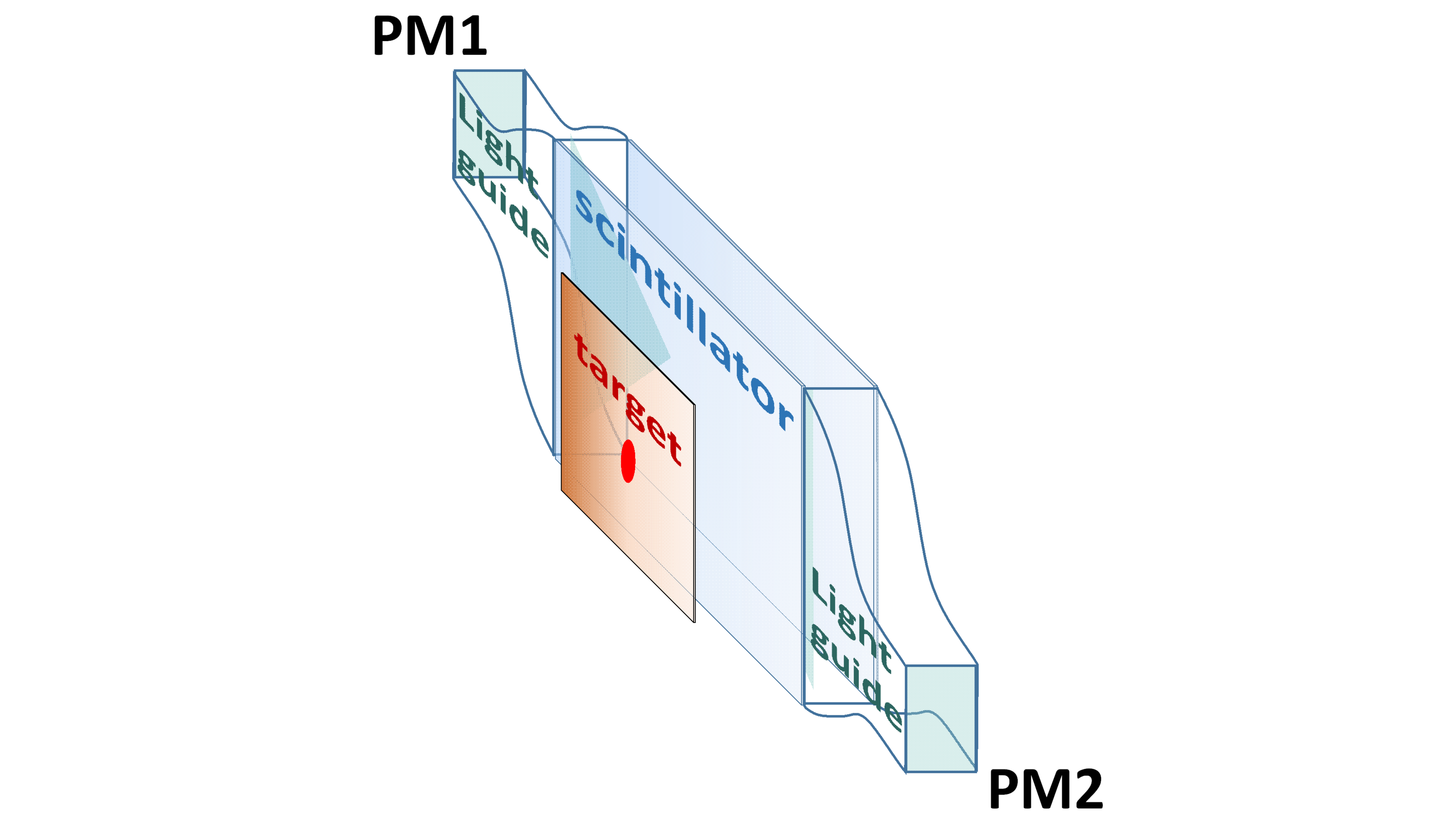}
\caption{\label{fig:detectors} Schematic view of the detection system (see text for details).}
\end{figure}
\par
\begin{figure}
\includegraphics[width=85mm]{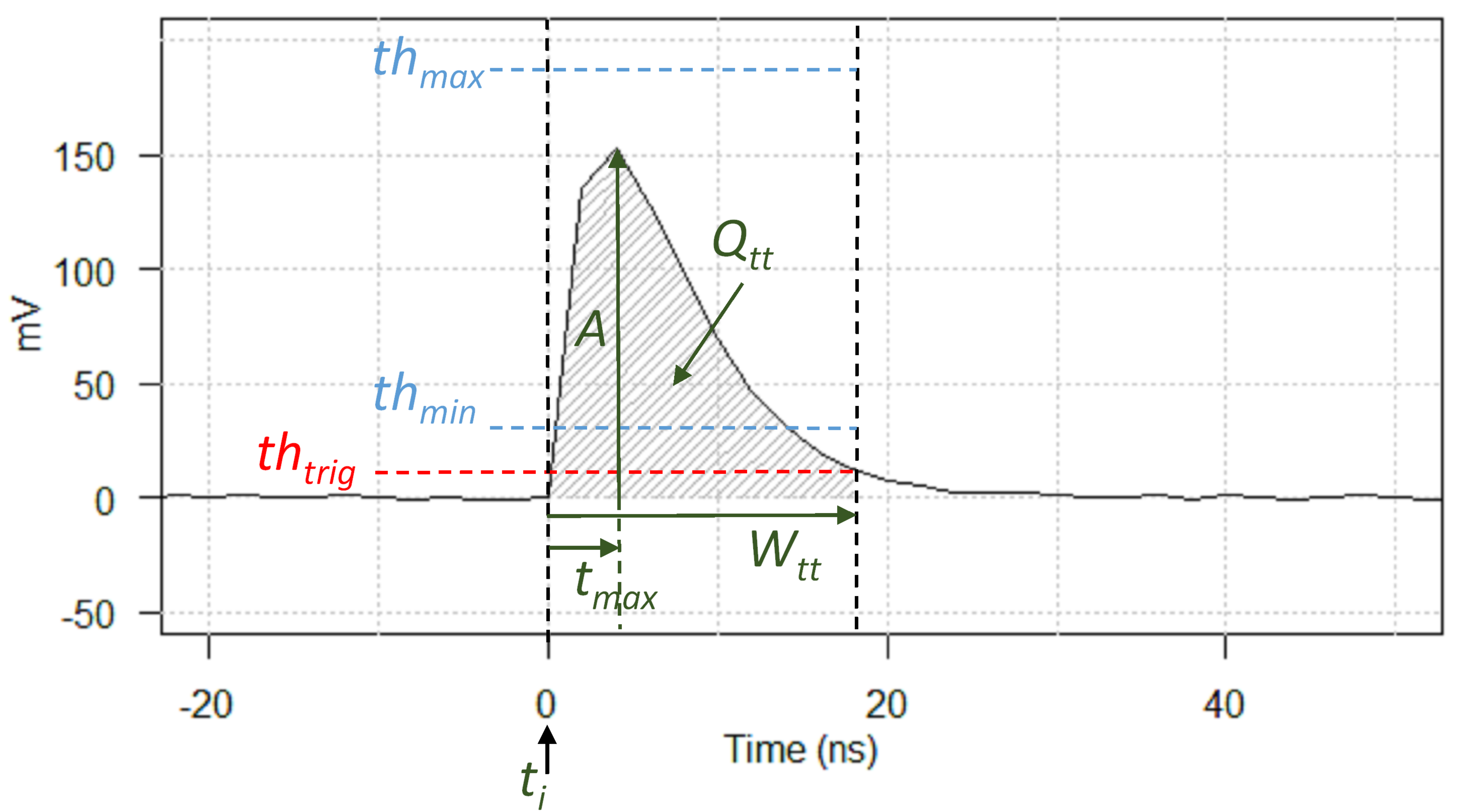}
\caption{\label{fig:MnMs} Illustration of the FASTER trigger logic and recorded parameters for the PM signals (see text for details).}
\end{figure}
The signals from the two PMs were directly sent to the real-time digital acquisition system FASTER~\cite{FasterWeb}. Three FASTER acquisition channels were used, each operating the digitization of the signals with a 500~MHz 12-bit converter with independent triggers based on individual thresholds. With this acquisition system, the digitized frames are processed in real time by FPGAs using predefined algorithms adapted to the measurements to be performed~\cite{FasterWeb}. All the data are time-stamped with a 2~ns step, allowing online and offline correlations over user-defined time windows. For the two PMs, we chose an algorithm processing the signals from threshold-to-threshold: as illustrated in figure 2, the digital sample to be processed is selected between the crossings of the signal rising and falling edges with a threshold level defined as $th_{trig}$. This choice yields samples of variable but optimal lengths with an additional dead time of only 2~ns between two successive digital data frames. For each selected frame, the algorithm provides: i) the time-stamp $t_i$ of the individual trigger with a 2~ns resolution, ii) the threshold-to-threshold integrated charge $Q_{tt}$, iii) the threshold-to-threshold duration $W_{tt}$, iv) the maximum amplitude $A$ within a 64~ns time window following $t_i$ and its date $t_{max}$. A second level trigger allows accepting or rejecting the events by selecting $A$ between the two additional thresholds $th_{min}$ and $th_{max}$. This second level trigger allows the use of a very low value of $th_{trig}$ without triggering on electronic noise. An illustration of the digitization performed by the system is given in figure 2. In addition, the baseline of the signals was continuously monitored for each channel and corrected for low-frequency variations (below 160~kHz) by the FASTER baseline restoration algorithm. A third channel using a specific RF algorithm of FASTER was dedicated to a comparison of the internal clock of the FASTER system with an independent OCXS oscillator~\cite{OCXS}, whose stability and accuracy are better than 0.1~ppm between -55 $^{\circ} $C and 85 $^{\circ}$C. The FASTER RF module~\cite{FasterWeb} was developed to perform precise measurements of the period or frequency of periodic signals such as, for instance, the frequency of a cyclotron. For each decay cycle, this module provided a measurement of the clock frequency difference with a relative precision of $9.5\times 10^{-9}$. The average relative deviation between the two clock frequencies was found to be $8.5\times10^{-7}$, with maximum variations remaining below $1.6\times 10^{-8}$ throughout the entire experiment. This deviation is thus negligible with respect to the level of precision aimed in the present work.
\par
Dead time due to the acquisition system can occur at three different stages. An intrinsic extendable dead time corresponding to the threshold to threshold signal duration plus 2~ns (one sample data point) is systematically present. Any other signal arising within this time window will cause pile-up and enlarge the threshold to threshold duration. The processing of the raw frames into calculated event data, such as displayed in figure 2, is done in real time and is not subject to losses. However, if the count rate is too high (typically larger than $10^6$ events per second), some of the data can potentially be lost as they can not be stored in the 2~kbytes buffers of the FPGA to be sent to the acquisition computer. This possible loss of data is constantly monitored using scalers counting independently both the calculated data and the data sent to the acquisition computer. The third possible source of data loss is due to the limited writing speed of the acquisition computer hard drive. It depends on the average data flux (count rate times the number of channels times the size of the processed data) and on the performance of the hard drive. When the writing speed of the latter saturates, then a complete buffer is lost. It is thus far the dominant cause of dead time that was observed. This possible loss of data is easily detected by comparing the number of data recorded on the hard drive to the counts of data provided by the scalers of the FPGA every 1~ms. Note that in the present experiment, a moderate detection rate was chosen such as to ensure no data losses at all besides pile-up.

\par
\section{Offline data analysis}
Compared to standard acquisition systems based on scalers, the advantages of a real-time digital multiparametric acquisition for half-life measurements of radioactive nuclei are evident. When relying on scalers, several runs using different thresholds and different hardware-imposed dead times are required to estimate the effect of pulse duration, of rate dependent gains, or of after-pulses~\cite{Triambak2012,Shidling2014}. In contrast, one can here use instead a unique set of data recorded with the lowest possible threshold (20~mV for the present work) and shortest dead time (threshold-to-threshold duration). The charge (or amplitude) of the signal and the time-stamp acquired for each event allow then an offline selection of the data by applying $\it{a~posteriori}$ the most appropriate dead time and threshold. Moreover, the charge information as a function of the time in the decay cycles and of the time interval between successive events provides means to carefully study the sources of systematic errors and to correct for remaining gain fluctuations and pile-up effects. Such careful studies, previously discussed for the analysis of the muon half-life measurement~\cite{Webber2011}, are mandatory when aiming at a relative precision better than $10^{-4}$. On the other hand, a multi-parametric acquisition system is somewhat slower than simple scalers, due to the larger pieces of information to be processed and recorded. For the present experiment, the maximum instantaneous acquisition rate was limited to about $2\times 10^5$~cps (counts per second), due to the writing speed of the acquisition computer hard drives. Nevertheless, in only 4 hours of data taking comprising 33 implantation-decay cycles, $\sim 3.5\times 10^8$ decays were recorded yielding a relative statistical uncertainty of $\sim10^{-4}$.
\subsection{Preliminary analysis and data selection}
A first analysis of the data from PM1 and PM2, coupled to the 5~mm thick scintillator, has shown an average duration of the signals $<W_{tt}>$ of 18~ns with a maximum duration of 44 ns. A non-extendable dead time of 50~ns was thus first applied to the data offline in order to to impose a common and constant dead time. Figures 3 and 4 show respectively the charge and time distributions obtained for PM1 after this selection during a complete decay cycle. On figure 3, the charge distribution of PM1 is compared to its counterpart (labeled PM1$_{coinc}$) when imposing a coincident signal from PM2 within a 16~ns time window (about twice the average rise time of the signals). This condition results in the disappearance of a large background contribution peaked at low charges. The constant-background charge distributions from PM1 signals obtained in the last 40~s of the cycle are also displayed for comparison. They were previously normalized to the complete cycle duration. The typical rate of the PM1 constant-background was about 12~s$^{-1}$ without coincidence and below 0.5~s$^{-1}$ in coincidence with PM2. In both cases, the constant-background contribution yields only signals with low charges.
\begin{figure}
\includegraphics[width=85mm]{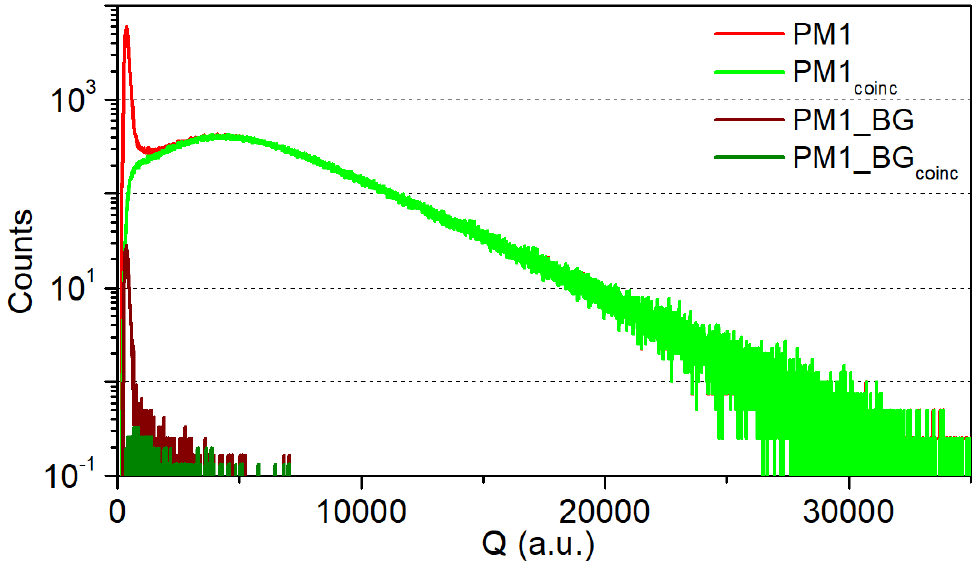}
\caption{\label{fig:Charge1} Charge spectra obtained for PM1 during one full decay cycle (the first cycle) without condition (red) and in coincidence with PM2 (green). Charge spectra obtained for the constant-background of the last 40~s and normalized to a full cycle without condition (wine) and in coincidence with PM2 (olive).}
\end{figure}
\begin{figure}
\includegraphics[width=85mm]{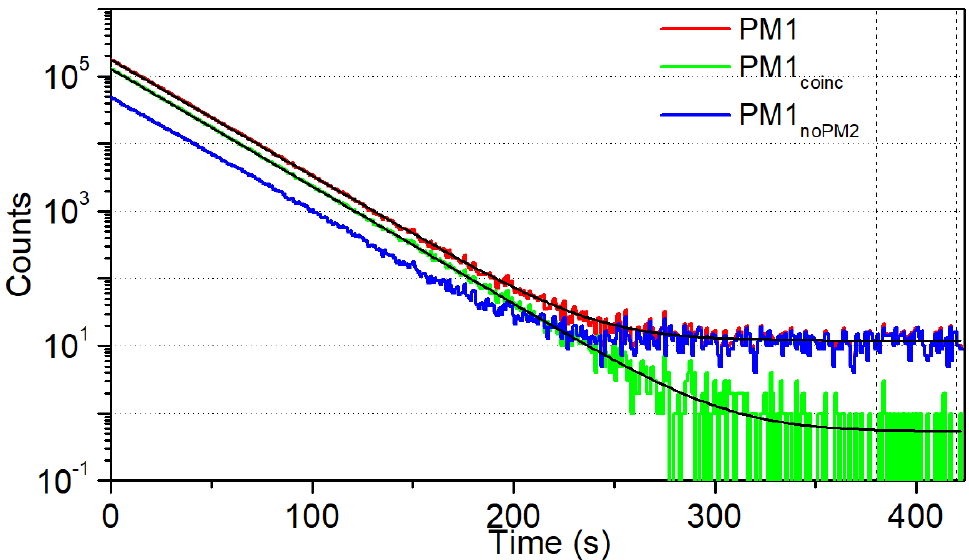}
\caption{\label{fig:Tcycle1} Decay time spectra obtained during one decay cycle (the first cycle) for PM1 without condition (red), in coincidence with PM2 (green) and with no coincidence signal on PM2 (blue). The bin width of the histogram is 1~s. The fit functions are plotted as black lines. The time window used for constant background selection is indicated by vertical dashed lines.}
\end{figure}
Figure 4 shows the time spectra obtained for PM1 signals, for PM1 signals detected in coincidence with PM2 (labeled PM1$_{coinc}$), and PM1 signals detected without triggering PM2 (labeled PM1$_{noPM2}$). The latter selection of events corresponds to the background peak at very low charge observed on the red curve of figure 3. For these background events, the time dependence shown on figure 4 is very similar to the one of events obtained in coincidence. This source of background strongly correlated to the decay rate was therefore attributed mostly to after-pulses and Cerenkov light emitted from the light guide or from the PM structure, and possibly to scattered or very low energy $\beta$ particles and photons triggering only one PM. The experimental data of figure 4 obtained for PM1 and PM1$_{coinc}$ were fitted using the decay function:
\begin{equation}
\label{Eq:Timestamp_distribution}
D_{neDT}(t)=\frac{D(t)}{1+neDT\cdot D(t)}
\end{equation}
with $D(t)=D_0 e^{(-t/\tau)}+BG$, $D_0$ being the initial decay rate, $\tau=T_{1/2}/\ln(2)$ the decay time constant, $BG$ the constant background rate, and $neDT$ the non-extendable dead time of 50~ns imposed by software. The fit method was based on likelihood maximization for data following the Poisson distribution.
In order to avoid any influence of the bin width chosen for the construction of the time decay spectra, the actual fit function applied to the histograms takes the form:
\begin{equation}
\label{Eq:Timestamp_distribution_int}
F_{neDT}(t_i)=\int_{t_i-\frac{\Delta t}{2}}^{t_i+\frac{\Delta t}{2}} D_{neDT}(t) \, \mathrm{d} t
\end{equation}
where $\Delta t$ is the bin width, $i$ is the bin number, and $t_i=i\Delta t+\frac{\Delta t}{2}$.
\par
At this stage of the analysis, the reduced $\chi^2$ of 671.0 for 421 degrees of freedom (424 bins of 1~s and 3 free parameters) obtained for the raw PM1 data, with a $P-value$ below $10^{-4}$, already indicates a disagreement with the model decay function. However, the reduced $\chi^2$ of 434.1 for PM1 data in coincidence with PM2 corresponds to a $P-value$ of 0.32, which does not allow questioning the model function or the data selection. As for experiments using a ``standard'' acquisition, it is now still required to look for an optimum threshold and dead time. For each cycle, the positions $Q_{max}$ of the maximum of the charge distributions of PM1 and PM2 were determined using a polynomial function of third degree. All the charge spectra were then normalized, as illustrated in figure 5(a), using the observable $Q_{norm}=Q_{tt}/Q_{max}$. This independent normalization for each cycle and the two PMs provides a common reference for applying a software threshold in charge $Q_{th}$. As done in figure 3, the figure 5(b) shows the contributions from constant background and from Cerenkov or after-pulses triggering only PM1. It already indicates that a minimum threshold close to $Q_{th}=0.25$ has to be applied on the charge $Q_{norm}$ after normalization. 
\begin{figure}
\includegraphics[width=85mm]{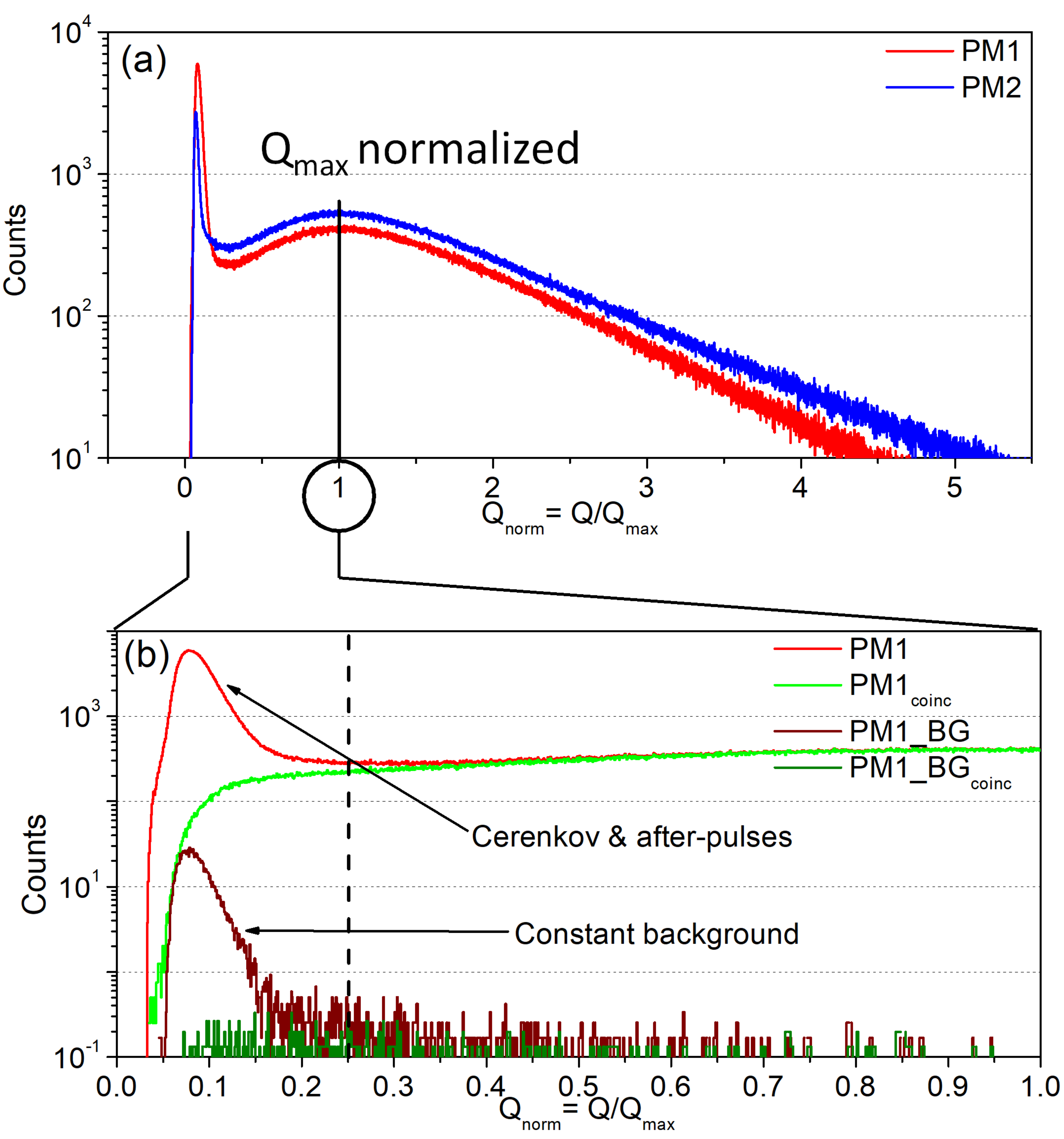}
\caption{\label{fig:Chargenorm} Charge spectra obtained during the first decay cycle for PM1 (red) and PM2 (blue) after normalization (a). The enlarged view (b) also shows the charge distribution for PM1$_{coinc}$ (green) and the constant background contributions (wine and olive). A threshold $Q_{th}=0.25$ is indicated by the vertical dashed line.}
\end{figure}
\par
An important feature of the acquisition system is the time-stamping of the data. By providing a precise time for each event, it is possible to look at the decay events as a function of the time-interval $\theta$ between successive triggers. The advantages of an analysis based on time-interval distributions were previously presented in \cite{Horvat2013}. Here, we use this information to optimize the data selection. Figure 6 shows the time-interval histograms obtained with PM1 for one cycle when applying different thresholds on the charge after normalization. They are compared to the theoretical distribution expected with no dead time \cite{Fontbonne2017}:
\begin{equation}
\begin{split}
\label{Eq:Time_Difference}
H(\theta)=\tau e^{-BG\cdot\theta} [\frac{e^{-D_e\theta}-e^{-D_0\theta}}{\theta} \\
+ BG\cdot Ei(-D_0\theta)-BG\cdot Ei(-D_e\theta)]
\end{split}
\end{equation}
where $Ei$ is the exponential integral function, $D_e= D_0 e^{(\frac{-T_e}{\tau})}$, $T_e$ is the duration of the cycle, and where the parameters $BG$, $D_0$, and $\tau$ are inferred from a fit of the decay time distribution. For PM1 data and a threshold $Q_{th}=0.15$, large deviations due to after-pulses are clearly visible, indicating a minimum appropriate offline dead time of $\sim1.0\ \mu$s. For PM1 data in coincidence with PM2 and higher thresholds, a deficit of counts below $1\ \mu$s is also observed. This can be attributed to other effects correlated with count rate such as pile-up, baseline and detector gain variations. Beyond these observations, the other obvious advantage of the time-stamp information is that an appropriate constant and precise dead time can be applied to the data afterward, by selecting events separated by a minimum difference in detection time. In the following analysis, we used a conservative software imposed dead time $neDT = 1.5~\mu s$ resulting in a negligible loss of statistics. This selection, combined to the use of eq. \ref{Eq:Timestamp_distribution} with the corresponding $neDT$ value, suppress the bias associated to after-pulses and account for missing data due to pile-up.
\begin{figure}
\includegraphics[width=85mm]{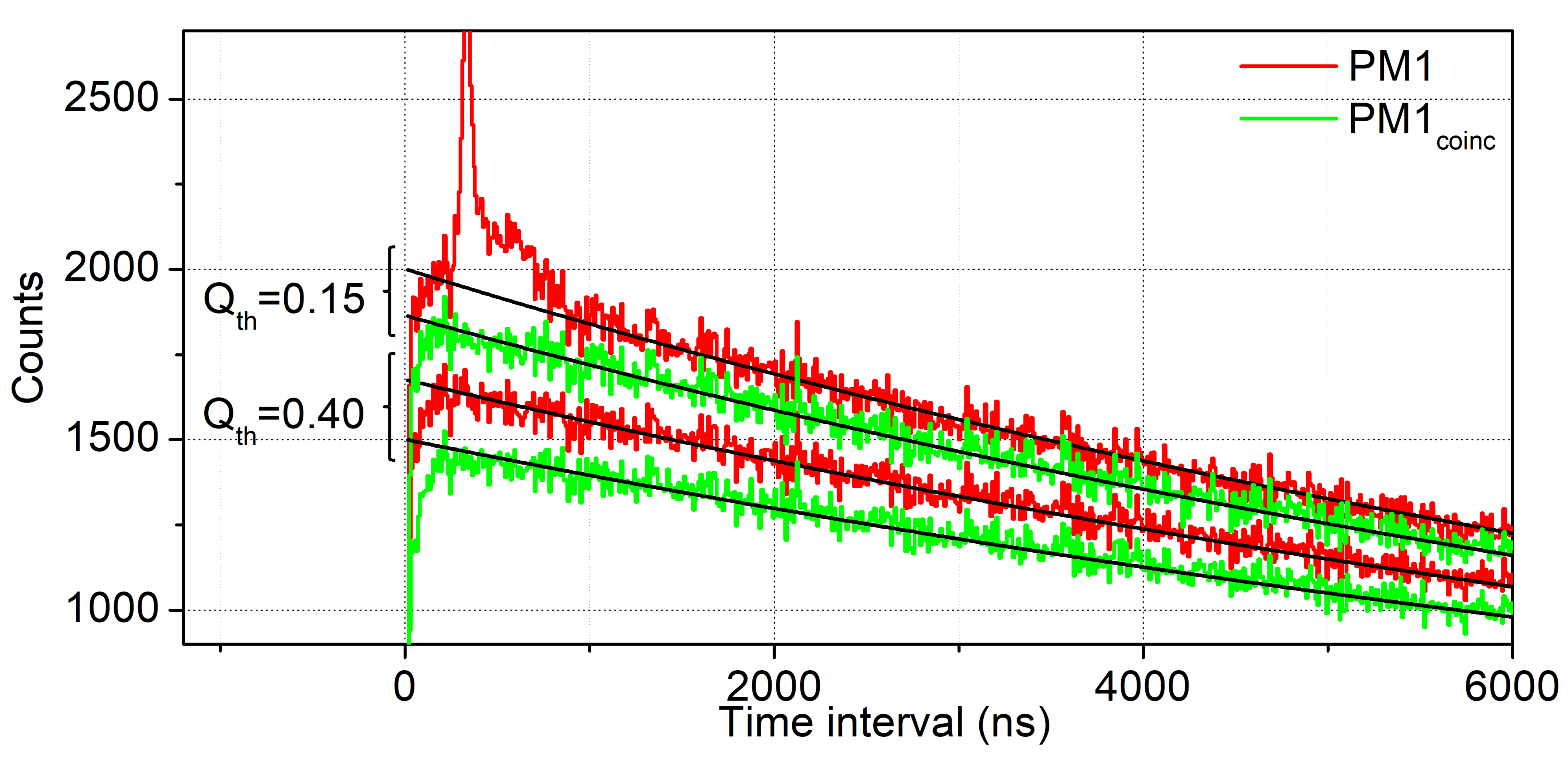}
\caption{\label{fig:Timeinterval} Time-interval spectra obtained for the first decay cycle with PM1 (red) and coincidences (green) using thresholds in charge $Q_{th}=0.15$ and $Q_{th}=0.40$. Theoretical distributions are given by black lines. The bin width of the histogram is 10~ns.}
\end{figure}
\subsection{Systematic effects}
After having estimated the appropriate minimum dead time and threshold, additional systematic effects such as gain variations, baseline variations, and second order effects due to pile-up were studied. During two short runs, each consisting of one decay cycle, the FASTER acquisition was used in an ``oscilloscope mode'' providing frames of 2.8~$\mu$s duration recorded at a 1~kHz rate. One run of 100~s duration was recorded with a high initial detection rate of $\sim 1$~Mcps and another run of 50~s duration with a moderate initial detection rate of $\sim $ 150~kcps, comparable to the experimental conditions for the 33 implantation-decay cycles. These oscilloscope frames were analyzed offline in order to estimate the baseline and the mean charge $<Q_{tt}>$ of the signals as a function of time within a cycle. Changes of $<Q_{tt}>$ with time were then interpreted as gain variations of the detection system. The results of this analysis for the four cases are displayed in figure 7. The baseline variations are found strongly correlated with the instantaneous decay rate. The gain variations are more difficult to interpret. For PM2 and for an initial detection rate of $\sim 1$~Mcps, it follows approximately an exponential growth with a time constant comparable to the nuclear decay one. For the three other cases, the gain oscillates without obvious pattern. For the $\sim $ 150~kcps initial rate, we found gain and baseline variations with amplitudes of respectively 0.9\% and 33~$\mu$V for PM1, and 2.3\% and 33~$\mu$V for PM2. For the high initial rate, it becomes 2.4\% and 558~$\mu$V for PM1, and 6.5\% and 685~$\mu$V for PM2. This indicates stronger variations with higher rates, and a better stability of PM1 compared to PM2. The variations of the baseline, similar for both PMs, could be interpreted as resulting from a small exponential decay tail in the PM signals each time a $\beta$ is detected. This tail has a typical initial amplitude of $V_{tail}\sim$100~$\mu$V and decays with a time constant of $\tau_{tail}\sim$5~$\mu$s that is too short to be properly corrected by the FASTER baseline restoration algorithm. The resulting baseline corresponds to the convolution of this exponential decay function with a Dirac comb of period inversely proportional to the decay rate:
\begin{equation}
\begin{split}
\label{Eq:BL_model}
BL(t,D_0)=V_{tail} \frac{e^{-\frac{1}{\tau_{tail}D_0 e^{(-t/\tau)}}}}{1-e^{-\frac{1}{\tau_{tail}D_0 e^{(-t/\tau)}}}}+V_0
\end{split}
\end{equation}
 where $V_0$ is the constant baseline at zero decay rate. The parameters $V_{tail}$, $\tau_{tail}$ and $V_0$ where extracted from a fit of the oscilloscope frames for PM1 and PM2. This modeling of the baseline variation was found in excellent agreement with the oscilloscope data. 
\par
\begin{figure}
\includegraphics[width=85mm]{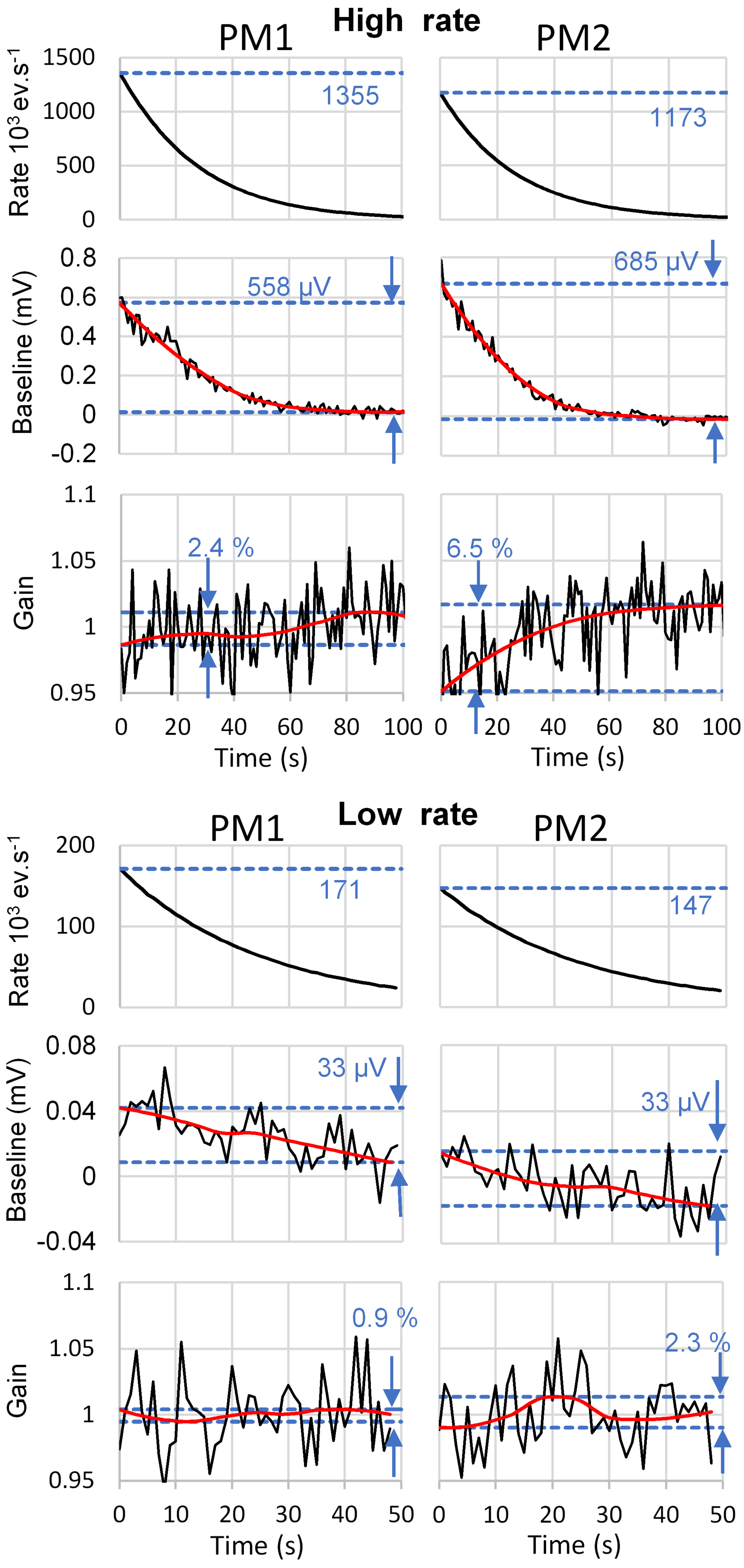}
\caption{\label{fig:Oscil_frames} Count rates, baselines variations, and gain variations inferred from oscilloscope frames for PM1 and PM2 at $\sim$1~Mcps and $\sim$150~kcps initial rates. Fluctuations due to the limited statistics are smoothened using a LOESS regression (LOcal regrESSion) from the R software package~\cite{Rweb}.}
\end{figure}
These effects, by changing the shape of the charge distribution as a function of time, modify the probability of accepting an event when applying a charge threshold and therefore affect the resulting decay time spectra. In a similar way, pile-up events whose probability is correlated with count rate modify the shape of the charge distribution with time. This second order effect is not taken into account in the dead time correction performed by eq.\ref{Eq:Timestamp_distribution}. The impact of these variations on the time decay spectrum obtained after applying a threshold in charge $Q_{th}$ has been thoroughly studied using a dedicated numerical model. The details of this work can be found in \cite{Fontbonne2017}. It will be the object of a separate publication and only the main ingredients are given in the following. The numerical data were generated using the signal duration $W_{tt}$ and charge $Q_{norm}$ density functions of the experimental events. These density functions were obtained after applying a non-extendable dead time $neDT = 1.5~\mu s$ to the data, by selecting decay events where pile-up is expected to be negligible, and after subtraction of the constant-background contribution extracted from the last 40s of each cycle.
In our model, gain and baseline variations were then applied numerically to the charge using an exponential time dependence. 
To simulate the effect of gain variation, the initial charge $Q_{norm}$ was multiplied by a factor $G(t)$ given by:
\begin{equation}
G(t)=1-\Delta G_0 e^{-\frac{t}{\tau}}
\label{Eq:gain_var}
\end{equation}
 where $\Delta G_0$ is the maximum gain variation. Note that in this study, we chose a gain variation model similar to the one observed for PM2 at $\sim$1~Mcps initial count rate.
Then, we added the charge offset due to the baseline fluctuation given by:
\begin{equation}
Q_{offset}(t)=BL(t,D_0) \times W_{tt}
\label{Eq:baseline_var}
\end{equation}
where $W_{tt}$ is the duration of the signal and $BL(t,D_0)$ the baseline modeled by eq.~\ref{Eq:BL_model}. 
The effect of pile-up on the recorded charge distribution was also accounted for by using a convolution of two charge distributions obtained when pile-up is negligible. The fraction of the charge spectrum resulting from pile-up events is then defined as:
\begin{equation}
p_e(i)=1-e^{-W_{tt}\frac{N_i}{\Delta t}}
\label{Eq:Pileup_fraction}
\end{equation}
where $N_i$ is the number of events detected in a given interval of time $\Delta t$. 
\par
Using this numerical model, a systematic study of the bias on the half-life estimation induced by gain and baseline variations as well as pile-up was then performed as a function of the threshold $Q_{th}$ chosen for data selection. Figure 8 shows the expected relative bias for PM1, resulting from gain variations with different values of $\Delta G_0$. Similar results, not shown here, were obtained for the signals of PM2. Even for small gain variations with $\Delta G_0=1\%$, we find a relative bias of 1.6~$\times 10^{-4}$ at the optimal threshold, close to $Q_{th}=0.2$. This bias increases almost linearly with $\Delta G_0$. It also increases strongly when moving $Q_{th}$ away from optimum. This shows that without an extremely stable PM gain (stability better than 1\%), a relative precision at the level of $10^{-4}$ requires gain monitoring and an appropriate selection of charge threshold. Note that for a gain variation of 6.5\%, as obtained with PM2 at high rates, the expected minimum relative bias is close to $10^{-3}$, far above the $10^{-4}$ statistical relative uncertainty of the present data.
\begin{figure}
\includegraphics[width=85mm]{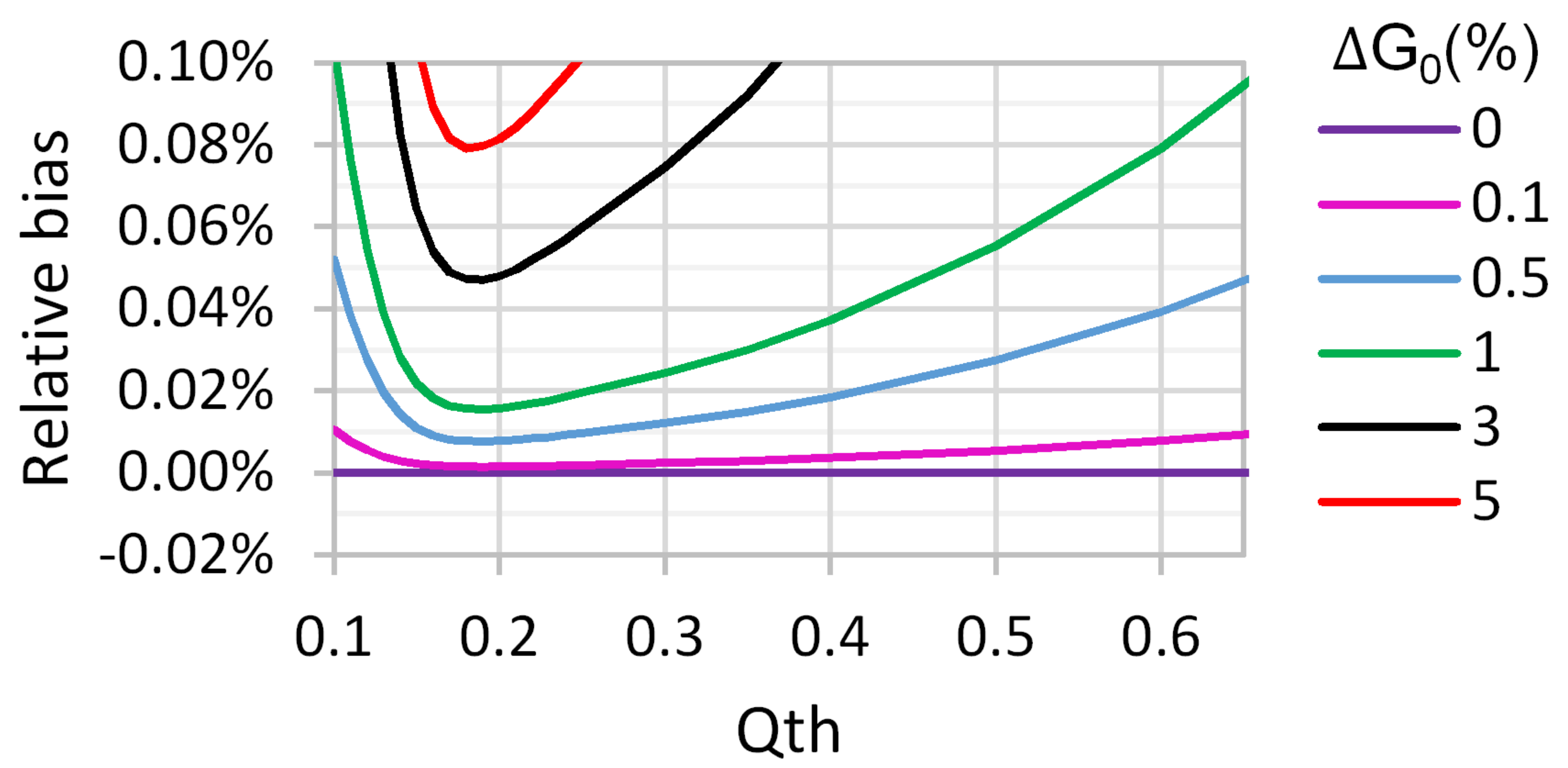}
\caption{\label{fig:Gain_bias} Expected relative bias on the half-life estimation for PM1 due to gain variations as a function of the threshold $Q_{th}$.}
\end{figure}
\begin{figure}
\includegraphics[width=85mm]{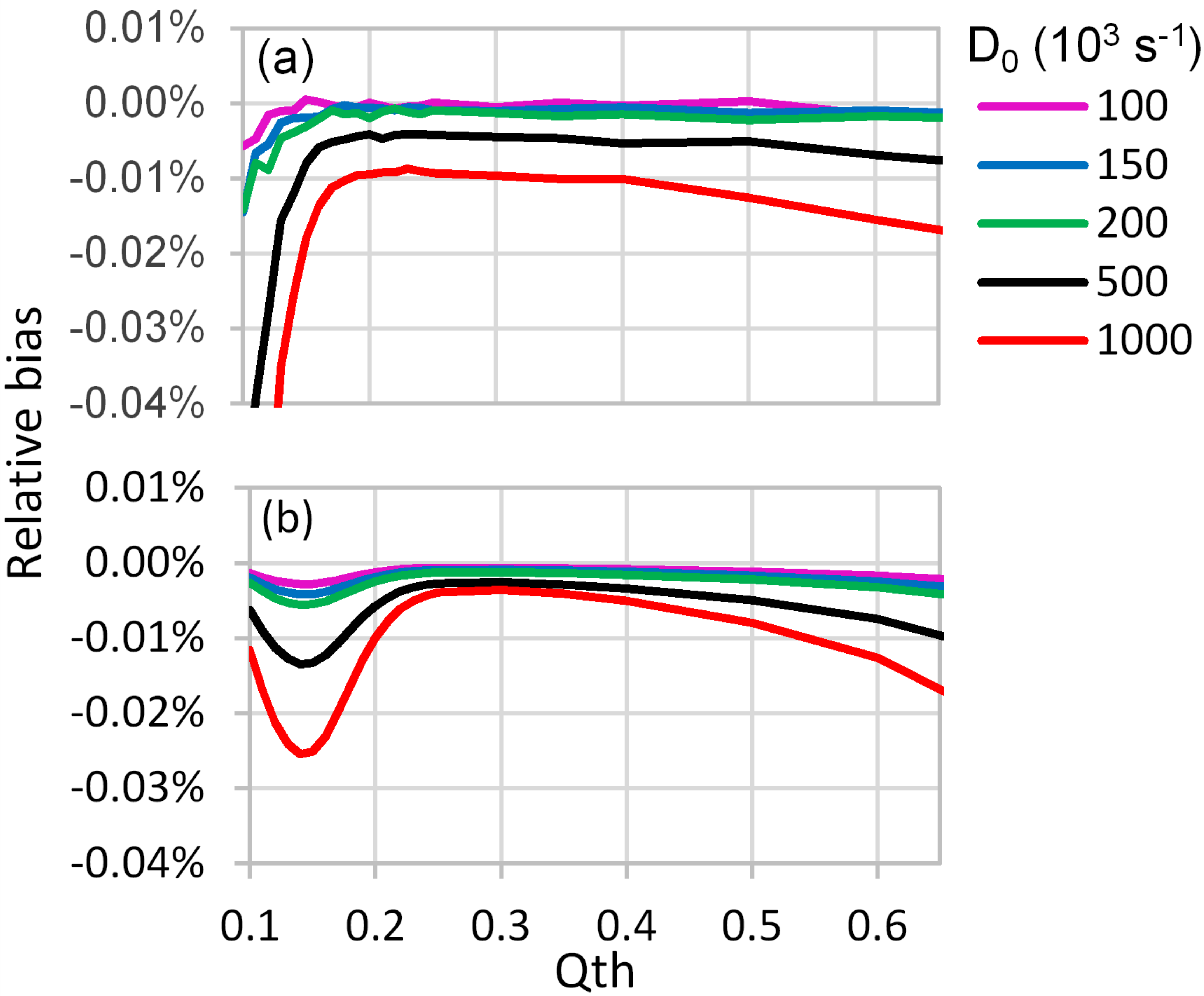}
\caption{\label{fig:Baseline_pileup} Expected relative bias on the half-life estimation for PM1 signals, due to (a) baseline variations and (b) pile-up as a function of the threshold $Q_{th}$ for different initial rates $D_0$.}
\end{figure}
Similarly, the effects of pile-up and of baseline variations for several initial detection rates and as a function of the threshold $Q_{th}$ are given in figure 9. As for the gain variation, a threshold in charge $Q_{th}$ close to 0.25 strongly reduces the bias. With this threshold and the detection rates of the present experiment (between 150~kcps and 200~kcps), the expected relative bias due to baseline variation is $-1.0\times 10^{-5}$ and the one due to pile-up $-0.6\times 10^{-5}$, with errors on these estimates below $10^{-6}$. However, running at a 1~Mcps initial rate would have induced a minimum relative bias of $\sim-10^{-4}$ due to baseline variation and of $\sim-3\times 10^{-5}$ due to pile-up.
From these results, it is clear that for a 1~Mcps initial detection rate the gain and baseline variations effects discussed above can cause systematic errors larger than $10^{-4}$. For the lower rate, the impact of pile-up and of baseline variations is negligible, but the relative bias induced by gain fluctuations as simulated in our model was found to be of the order of $10^{-4}$.
\par
To confront the model to real data, we applied the same procedure using the gain variations of figure 7 obtained during the unique cycle dedicated to oscilloscope frames and recorded with a $\sim$150~kcps initial detection rate. The half-lives as a function of $Q_{th}$ given by the model with PM1 and PM2 charge distributions are shown in figure 10. It assumes a non-biased initial half-life of 17.2569~s, which is the final result of our analysis. It is compared to the experimental results given by the weighted average of the 33 time decay values, obtained by fitting the decay time spectra of PM1 and PM2 data with eq. \ref{Eq:Timestamp_distribution_int}. The data were filtered beforehand by applying a non-extendable dead time of 1.5~$\mu$s and the thresholds $Q_{th}$ indicated on the horizontal axis. The half-lives obtained for data corresponding to coincidences between PM1 and PM2 are also shown.
\begin{figure}
\includegraphics[width=85mm]{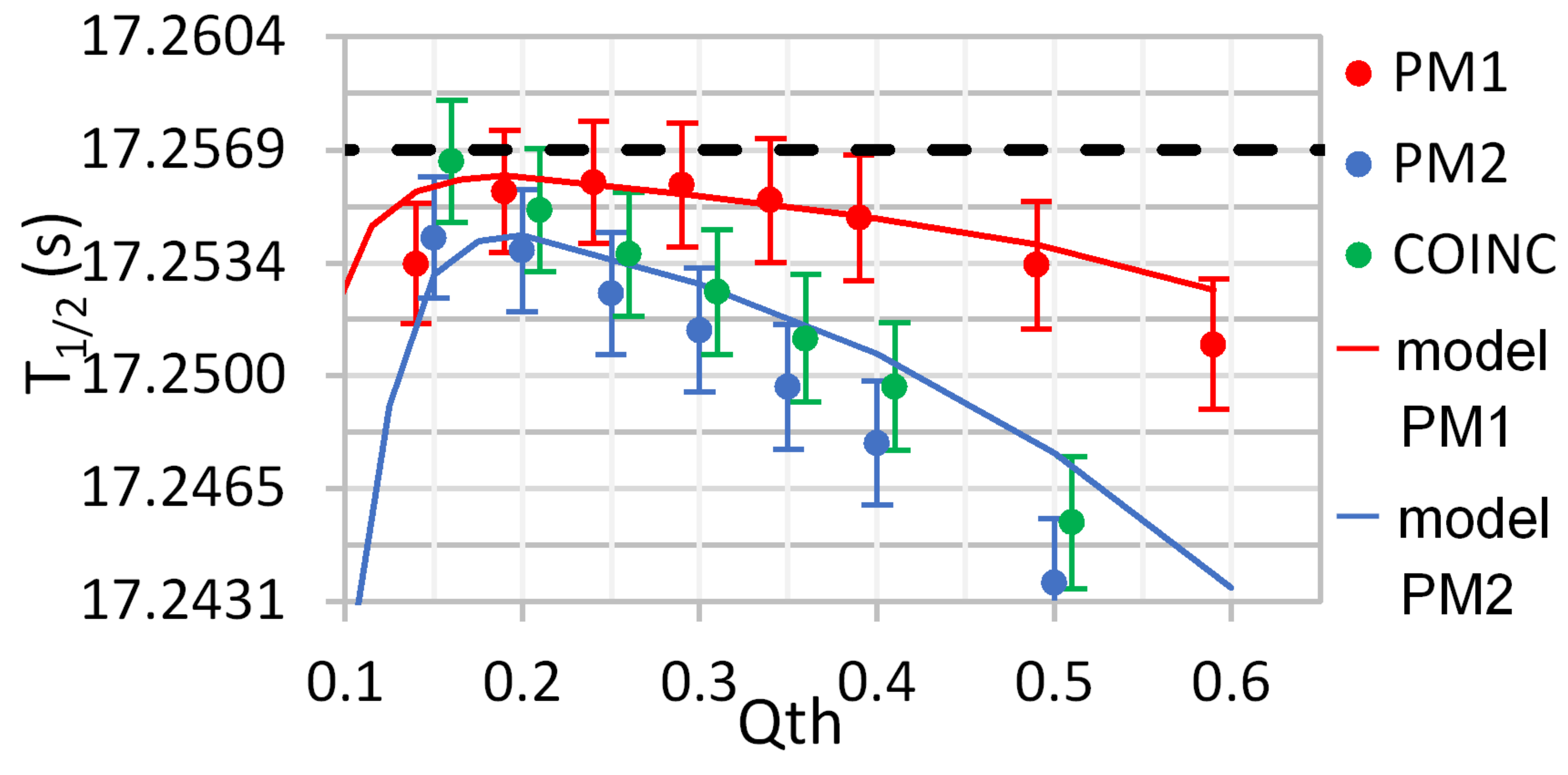}
\caption{\label{fig:Test_model} Half-lives as a function of $Q_{th}$ given by the model with PM1 and PM2 charge distributions using gain variations from figure 7 (lines), compared to half lives resulting from the fit of the 33 time decay curves obtained with PM1, PM2, and coincidence experimental data (points). The horizontal dashed line indicates the final value of the half-life resulting from the present analysis (see text for details).}
\end{figure}
Despite the fact that the gain fluctuations occurring during the 33 decay cycles may have been different than the gain fluctuations measured with our single oscilloscope run, we observe a good agreement between the model and the experimental data. Minimum relative bias of 5$\times10^{-5}$ and 1.2$\times10^{-4}$, for PM1 and PM2 respectively, are obtained for $Q_{th}$ close to 0.25 for PM1 and 0.2 for PM2. The dependence of the bias to $Q_{th}$ follows a pattern similar to the one of figure 8, but with opposite sign. This is due to the shape of the gain variations obtained with the oscilloscope data at low rate that differs from the exponential growth previously considered in the model. One can also note that selecting PM signals in coincidence does not reduce the bias. On the opposite, figure 10 shows that for most values of $Q_{th}$, the selection of coincidence events results in a bias close to the one obtained for PM2 data (more affected by gain fluctuations) that is larger than for PM1 data alone.
\subsection{Event by event compensation and final results}
In spite of the apparent good agreement obtained when using the oscilloscope data to model the effect of gain fluctuations, a more robust analysis requires access to gain fluctuations occurring along each decay cycle. For the final analysis of the data, a gain compensation algorithm was thus developed. The decay time spectrum of each cycle was first divided into time-intervals containing $5\times10^3$ events. Using the mean charge recorded during these intervals and a recursive procedure~\cite{Fontbonne2017}, the gain variation from one interval to another was inferred. In this procedure, the negligible effect of the $\sim30\ \mu$V baseline variation was not accounted for. The gain variations were then compensated on an event by event basis, prior to applying the charge threshold. The decay time spectra were eventually fitted using eq. \ref{Eq:Timestamp_distribution_int} with a non-extendable dead time of 1.5~$\mu$s. Three different minimization methods were used: the Newton method from the R statistical computing package, the Levenberg-Marquardt algorithm, and the quasi-Newton method from MIGRAD, all yielding the same result. The half-lives obtained that way for PM1, PM2, and coincidence events with different thresholds in charge $Q_{th}$ are given in figure 11.
\begin{figure}
\includegraphics[width=85mm]{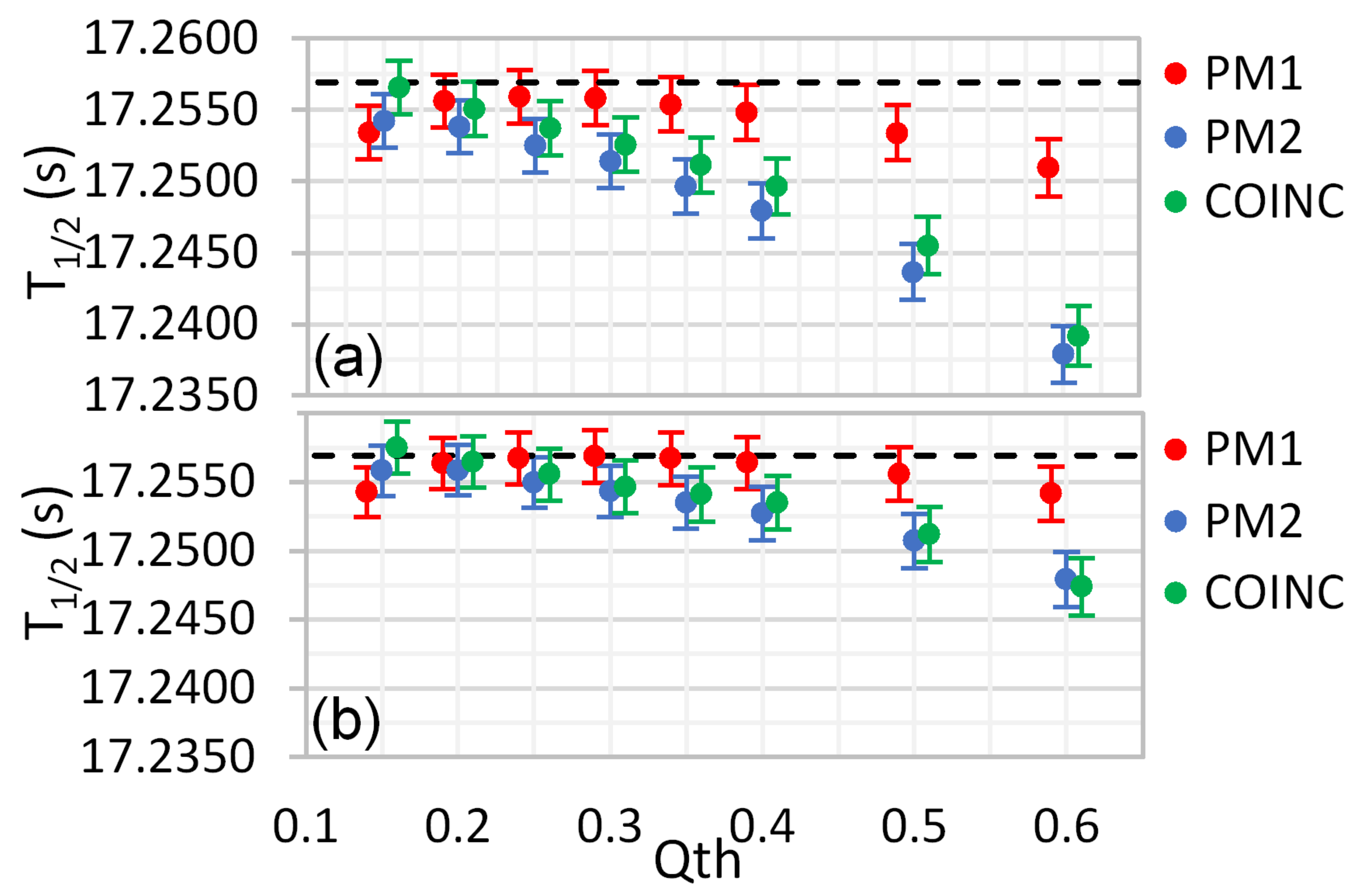}
\caption{\label{fig:Comp_results} Half-lives as a function of $Q_{th}$ obtained for PM1, PM2 and coincidences before (a) and after (b) applying the gain compensation algorithm. The horizontal dashed lines indicate the final estimation of the half-life based on the present analysis (see text for details).}
\end{figure}
After compensation for gain variations, the half-life estimations obtained with PM1 become very weakly dependent on the threshold $Q_{th}$ and remain constant from $Q_{th}=0.2$ to $Q_{th}=0.4$. For PM2 data, more affected by gain variations, the gain compensation effect is less effective. The half-life estimations obtained for the optimal threshold, close to $Q_{th}=0.2$, are in good agreement with PM1 data but the compensation does not completely correct for the deviations observed with higher thresholds. This limitation was attributed to the available statistics when measuring gain fluctuations along the cycles. The procedure allows to measure gain variations with a statistical uncertainty still better than the correction itself only up to 140~s in the decay cycle. For PM2, gain fluctuations beyond 140~s within the cycles, not accessible here, still impact the data. As a consequence, we only kept the result obtained with PM1 at the optimum threshold $Q_{th}=0.25$ yielding the estimate T$_{1/2}=17.2569\pm0.0019_{(stat)}$~s after including the small aforementioned corrections for pile-up and baseline variations. 
\par
The partial failure of the gain compensation algorithm for PM2 motivated further studies to estimate potential systematic errors induced by the compensation. This was done by the mean of Monte Carlo (MC) simulations of 100~cycles with statistics comparable to the experimental data. For the time distribution, we used eq. \ref{Eq:Timestamp_distribution} with a non-extendable dead time of 1.5~$\mu$s, T$_{1/2}=17.2569$~s, and values of $D_0$ and $BG$ equal to the ones inferred from experimental data. For the charge distribution, the experimental data of PM1 signals were directly used. The simulations were performed without gain variation along the cycles duration. Then, the procedure to measure and correct gain variations was applied to the MC data. The relative error on the gain measurement due to the limited statistics remained below 0.6\% and resulted in a systematic error on the half-life estimation $\Delta$T$_{1/2(syst)}$=0.0009~s.
\begin{figure}
\includegraphics[width=85mm]{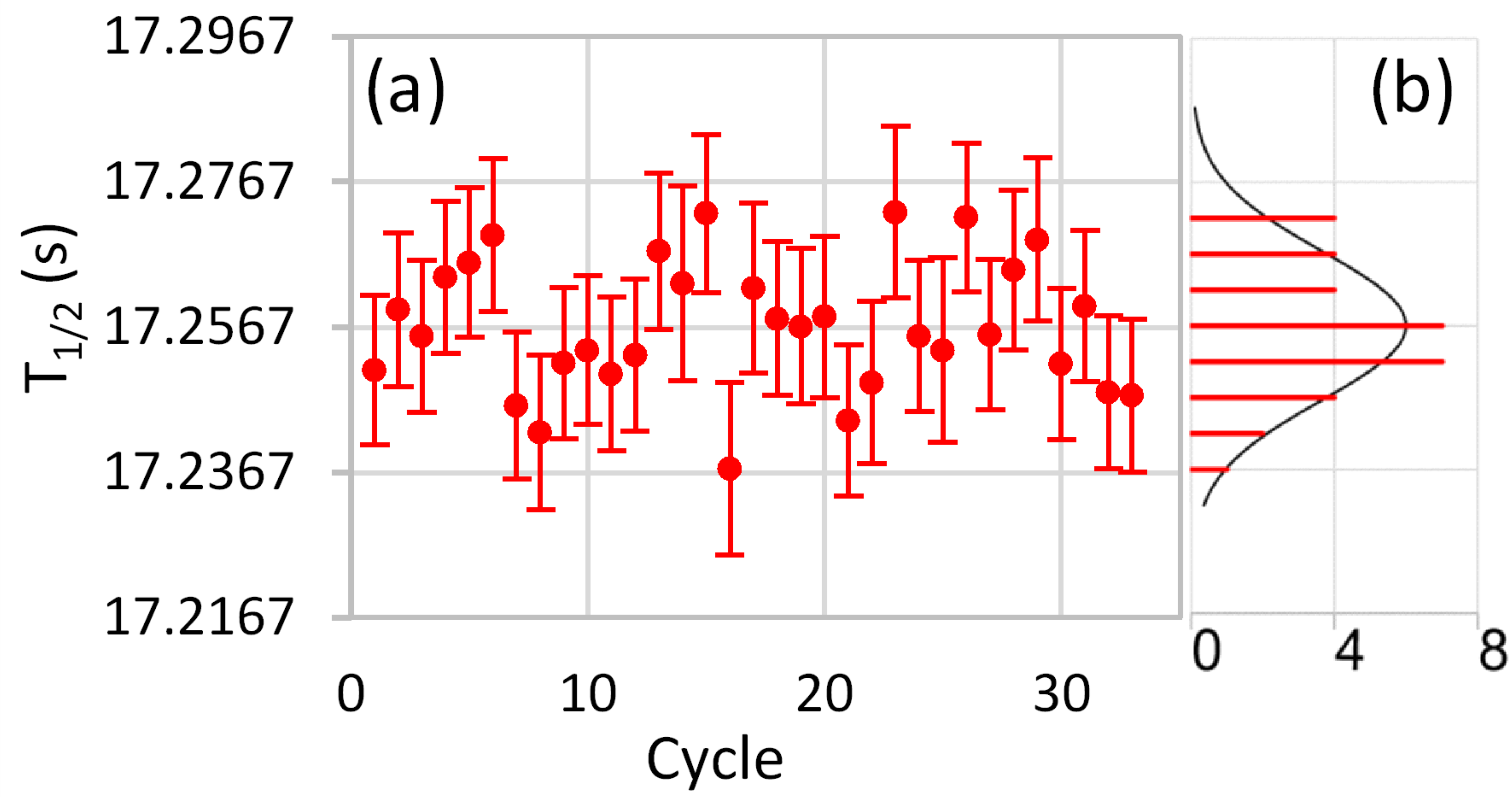}
\caption{\label{fig:Comp_cycles} Half-life estimations with PM1 for the 33 cycles using $Q_{th}=0.25$ gain compensation (a). Normalized residuals distribution (b).}
\end{figure}
\par
 Software applied dead times ranging from 1~$\mu$s to 4~$\mu$s were also investigated, showing no significant effect besides the additional loss of statistics. Similarly, the dependence on initial count rate was also studied by changing the lower limit of the fit applied to the time decay spectra. Again, no notable variation of the half-life estimation resulted from these tests. As shown on figure 12, the half-lives obtained by fitting individually each cycle were statistically consistent, without any trend indicating a slow drift along the complete duration of the experiment. The $\chi^2$ distribution obtained for PM1 with the 33 cycles, shown in figure 13, is perfectly compatible with the one expected for 421 degrees of freedom.
\begin{figure}
\includegraphics[width=85mm]{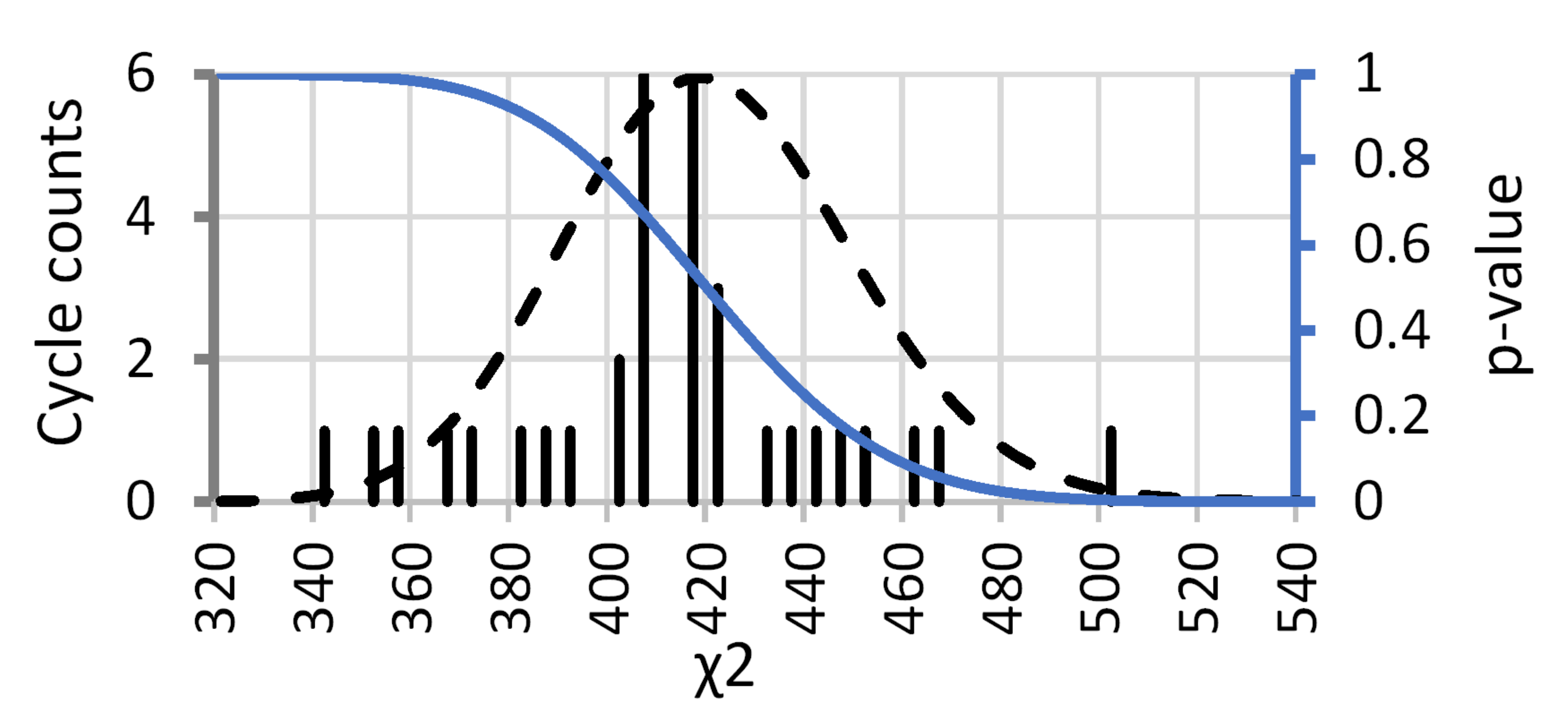}
\caption{\label{fig:chisquare1} Chi square distribution of the 33 cycles using PM1 data, $Q_{th}=0.25$, and gain compensation. The associated $P-values$ and the shape of the theoretical $\chi^2$ distribution (black dashed line) are also indicated for 421 degrees of freedom.}
\end{figure}
\par
Our final estimation of the half-life of $^{19}$Ne, T$_{1/2}=17.2569\pm0.0019_{(stat)}\pm0.0009_{(syst)}$~s, is found in perfect agreement with the former measurements of Triamback {\it et al.}~\cite{Triambak2012} and Ujic {\it et al.}~\cite{Ujic2013}, as illustrated in figure 14. However, it differs by 3.2 standard deviations from the value obtained by Broussard {\it et al.}~\cite{Broussard2014} and tends to discard the latter. Despite the rigorousness of the Broussard and collaborators analysis, the total contribution of the numerous corrections, all at the $10^{-4}$ level, that were applied (for dead time, accidental coincidences, diffusion, background, and contamination) may have been underestimated.
\begin{figure}
\includegraphics[width=85mm]{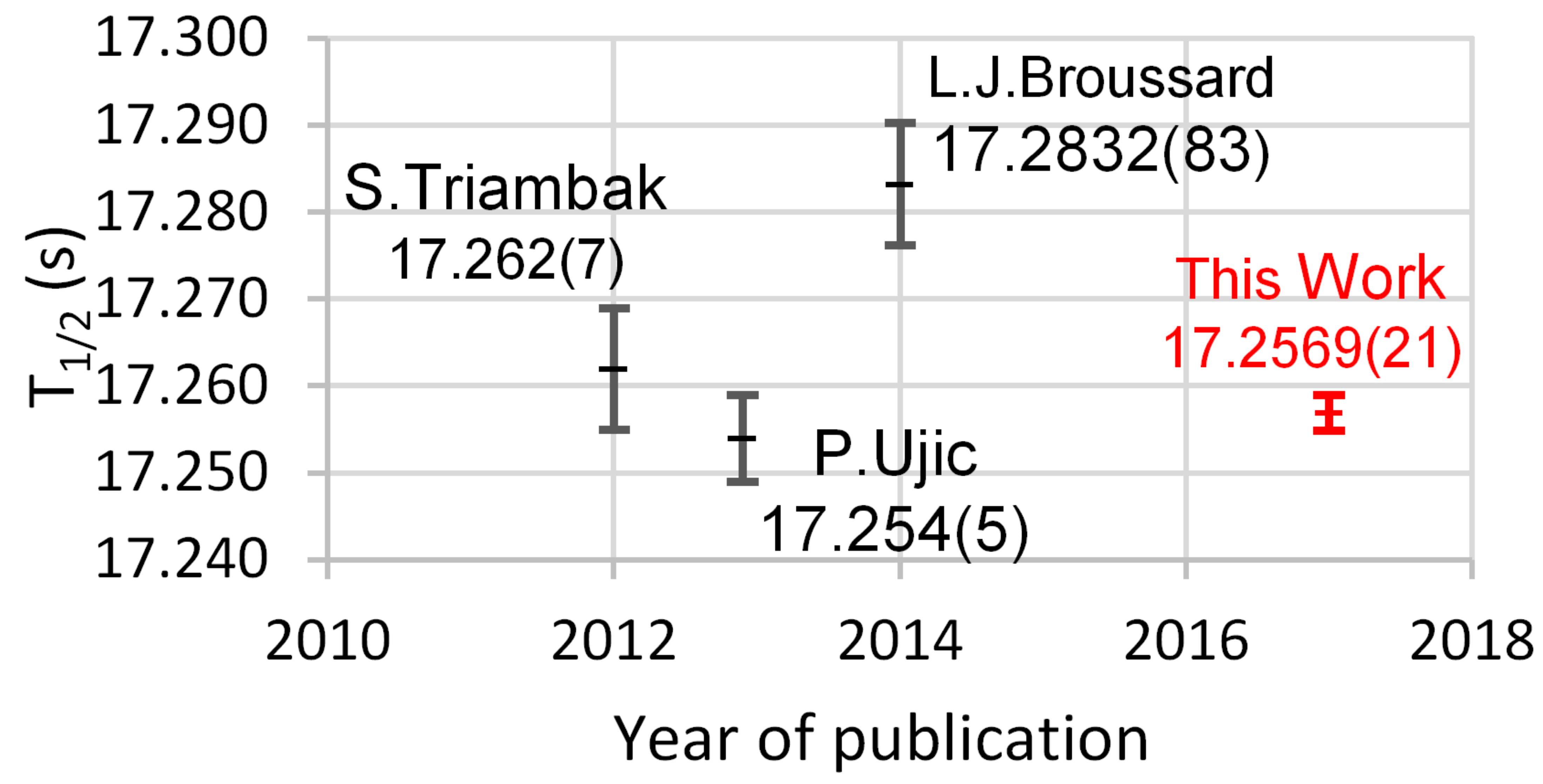}
\caption{\label{fig:Ne_halflife} Most recent and precise measurements of $^{19}$Ne half-life.}
\end{figure}
\subsection{Warnings}
We have shown in the previous section that gain variations, baseline variations, and pile-up can affect a decay time spectrum and result in a significant bias for half-life estimations. When aiming at a relative precision of $10^{-4}$ or below, pile-up and baseline fluctuations seem to remain negligible for the moderate initial count rates chosen for the present work but may become significant for rates of $\sim$1~Mcps and higher. The impact of gain fluctuations was found more critical, as even for small variation of the order of 1\% and an optimal threshold, a relative bias of the order of $\sim10^{-4}$ can be expected. To detect such a bias without dedicated measurements, one can only rely on a verification of the $\chi^2$ distribution and hope that it is sensitive to this systematic effect. As shown in figure 11(a), the half-life estimation obtained for PM2 without gain compensation when applying a threshold $Q_{th}=0.6$ is about 9~standard deviations away from our final result. Figure 15 shows the associated $\chi^2$ distribution for the 33 cycles. It is still perfectly compatible with what one would expect for 421 degrees of freedom. Similar $\chi^2$ distributions (not shown here) were obtained for all the results given by figure 11(a), despite large disagreements between these biased half-life estimations and the final one. We can thus conclude that with the present statistics yielding relative statistical uncertainties of $\sim10^{-4}$, the $\chi^2$ test does not allow to detect a relative bias due to gain variation smaller than $\sim10^{-3}$. By contrast, the systematic study of the decay spectrum as a function of charge threshold performed in the present analysis was found to be very sensitive to such sources of bias.
\begin{figure}
\includegraphics[width=85mm]{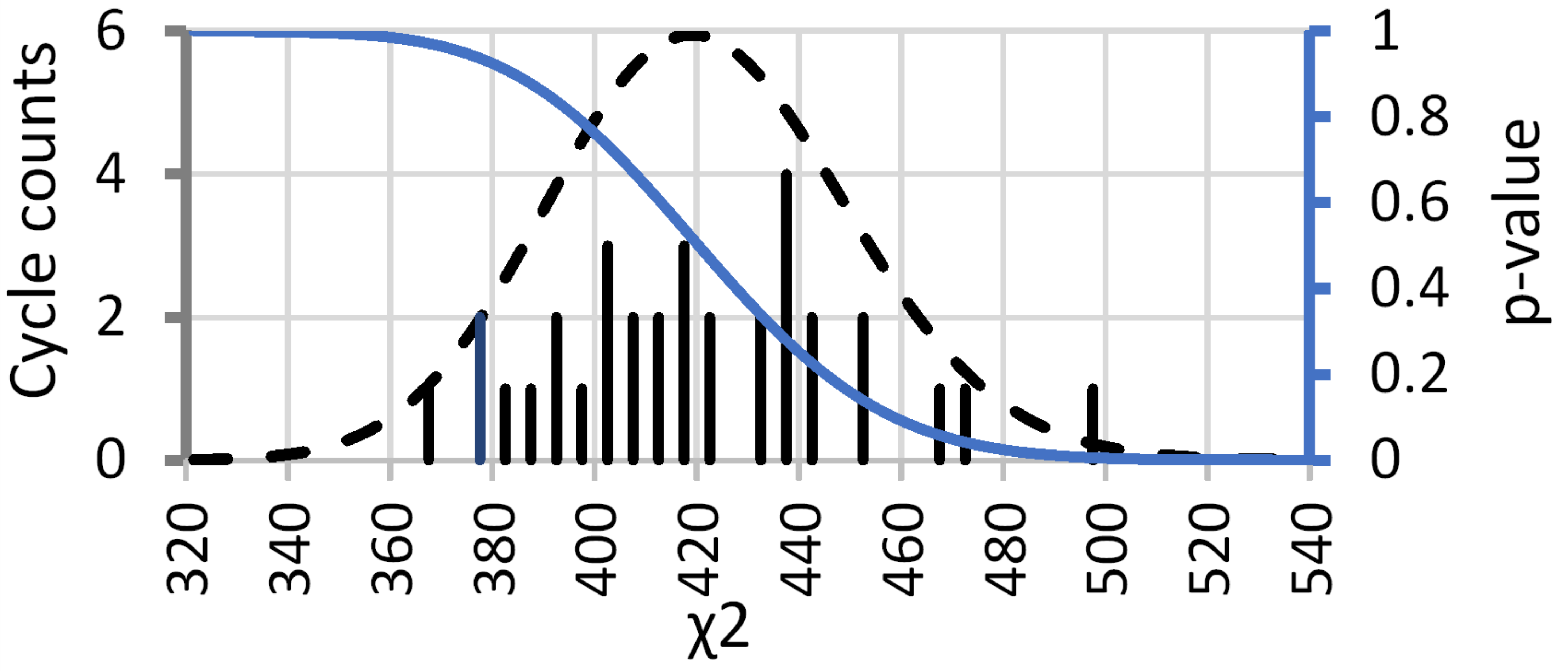}
\caption{\label{fig:chisquare2} Chi square distribution of the 33 cycles using PM2 data, $Q_{th}=0.6$, and no gain compensation. The associated $P-values$ and the shape of the theoretical $\chi^2$ distribution (black dashed line) are also indicated for 421 degrees of freedom.}
\end{figure}
\section{Future improvements}
To improve the relative precision of $\sim10^{-4}$ obtained with the present measurement, several simple upgrades can be foreseen. The main limitations were here due to statistics. With two days of beamtime (instead of only four hours) and a second detector to double the solid angle, a gain of a factor $\sim\times 24$ can readily be obtained. The initial event rate per channel can be easily increased by a factor of two without data losses in the acquisition by using two hard drives with higher writing speed. Other minor improvements can also be performed by optimizing the cycle duration (gain of$\sim\times 1.2$) and by limiting the Cerenkov contribution in the light guides (gain of$\sim\times 1.3$). The overall gain in statistics would lead to a relative statistical error of $\sim10^{-5}$. Note that such a statistics is accessible in a reasonable time only by using fast detectors, such as scintillators, and with beam intensities of the order of a few $10^{5}$ pps. 
\par
At this level of precision, pile-up, baseline and gain fluctuations have to be considered very carefully. The bias due to pile-up, on the order of $\sim10^{-5}$, can be estimated with a relative precision better than $10^{-6}$ thanks to the knowledge of the $Q_{tt}$ and $W_{tt}$ distributions. The effect of baseline variations was also precisely estimated in the present work, thanks to the oscilloscope data and a variation only correlated to the detection rate. One could however benefit from a measurement of the baseline for each detected event by using an additional charge integration window just before the trigger. The FASTER system readily allows the use of such a window within the 20~ns preceding the trigger and this feature will be implemented in the future versions of the acquisition.
\par
The gain variation is more problematic as i) the associated minimum bias was at the level of $10^{-4}$ for the present experiment and ii) the relative error on the gain measurement was limited to 0.6\%, yielding a relative systematic error of $5\times10^{-5}$ on the half-life estimate. A promising improvement consists in using a monochromatic calibration source during the half life measurement. Figure 16 shows the expected charge spectrum obtained in one second with a scintillator irradiated by $2.5\times10^5$ beta particles from $^{19}$Ne decay and alpha particles from a 2~kBq $^{241}$Am source. Due to quenching, the light produced by the 5.5~MeV alpha particles corresponds to the light emitted by $\sim350$~keV beta particles. Considering a 10\% energy resolution at 1~MeV for beta particles and a $^{241}$Am activity of 20~kBq, the expected precision on gain monitoring for one second using the mean position of the alpha peak is 0.07\%. This would reduce the relative error due to gain variations on the half-life estimate below $7\times10^{-6}$. Another approach would consist in limiting the intrinsic gain variation of the detection setup. For this purpose, it is now also foreseen to test the replacement of the PMs by Silicon photomultipliers, whose stability could be better.
\begin{figure}
\includegraphics[width=85mm]{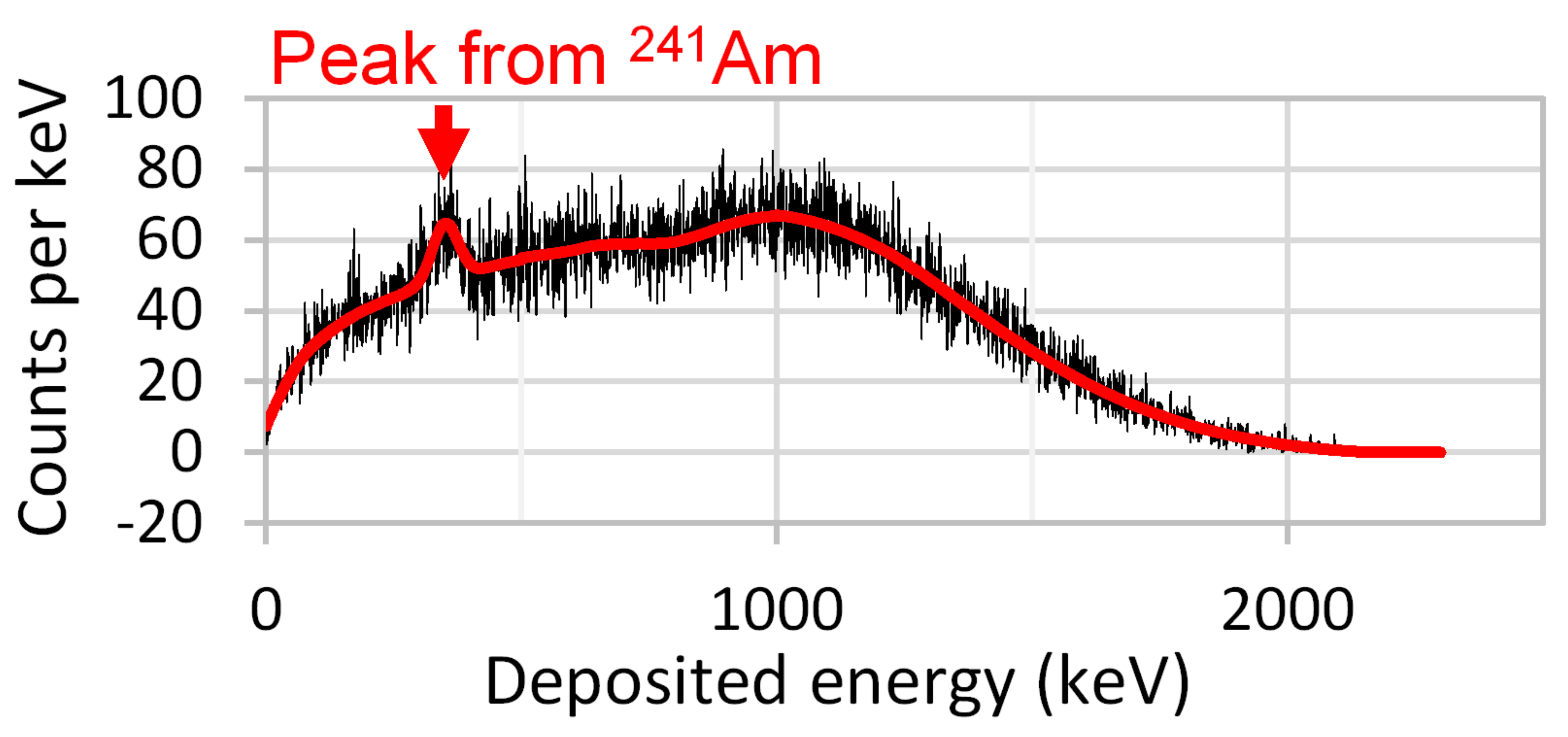}
\caption{\label{fig:Am241} Simulation of the deposited energy spectrum in a 5~mm thick BC400 scintillator irradiated by $2.5\times10^5$ beta particles from $^{19}$Ne decay and $10^3$ alpha particles from $^{241}$Am. The horizontal axis indicates the energy in keV electron equivalent.}
\end{figure}
\section{Conclusion}
Thanks to a real-time digital multiparametric acquisition system, the half-life of $^{19}$Ne has been measured with the unprecedented relative precision of $1.2\times10^{-4}$. The result, in agreement with the measurements of Triambak et al. and Ujic et al. ~\cite{Triambak2012,Ujic2013}, is not consistent with the last measurement by Broussard et al. \cite{Broussard2014}. The detailed analysis of the data provided by the multiparametric acquisition system has shown that at this level of precision, pile-up, baseline variations and primarily gain variations have to be considered carefully. Recording additional parameters such as integrated charge and absolute time-stamp allowed to both, determine the optimal dead time and threshold, and to check the conformity of the model decay function. It was also demonstrated that a study of the decay time dependence on the charge threshold is very sensitive to bias that can not be revealed by a $\chi^2$ test. Several possible improvements have been discussed, showing that relative precisions of the order of $10^{-5}$ can be achieved for half-life measurements of nuclei with production rates above $10^5$~pps.

\begin{acknowledgments}
We would like to express our gratitude to the GANIL crew for delivering the radioactive beams and for friendly collaboration. We acknowledge the support of European ENSAR project cycles (No. 262010), from French-Romanian Collaboration Agreement IN2P3-IFIN-HH Bucharest No. 03-33, the R\'egion of Basse Normandie, Helmholtz Association (HGF) through the Nuclear Astrophysics Virtual Institute (NAVI) and LEA IN2P3-ASCR NuAG projects and Ministry of Education, Science and Technological Development of Serbia - The project P171018. We thank Oscar Naviliat-Cuncic for useful discussions.
\end{acknowledgments}

\end{document}